\documentclass[twocolumn,prd,nofootinbib]{revtex4-2}

\usepackage{geometry}
\usepackage{graphicx}
\usepackage[caption=false]{subfig}

\usepackage{amsmath, bm}
\usepackage{hhline}
\usepackage{xcolor}

\usepackage{verbatim}
\usepackage{dcolumn}
\usepackage{color}
\usepackage{xspace}
\usepackage{url}
\usepackage{float}
\usepackage{multirow}
\usepackage{hyperref}

\newcommand{\secref}[1]{Section~\ref{#1}}
\newcommand{\figref}[1]{Figure~\ref{#1}}
\newcommand{\tabref}[1]{Table~\ref{#1}}
\newcommand{\eqnref}[1]{\eqref{#1}}                       
\newcommand{\Secref}[1]{Section~\ref{#1}}

\def\apj{Astrophysical Journal}                %
\def\apjl{Astrophysical Journal}             %

\def\prd{Phys.~Rev.~D}       %
\def\cqg{Class.~Quant.~Grav.~}%

\newcommand{\Lmarg}{\mathcal{L}_{\rm{marg}}(\pmb{\lambda})}
\newcommand{\tfe}{\texttt{TaylorF2Ecc}}
\newcommand{\teo}{\texttt{TEOBResumS}}
\newcommand{\mc}{{\cal M}_{\rm{ch}}}
\newcommand{\mcecc}{{\cal M}_{\rm{ch}}^{\rm{ecc}}}

\def\RIT{Center for Computational Relativity and Gravitation, Rochester Institute of Technology, Rochester, New York 14623, USA}

\begin{document}

\title{Parameter Estimation for Low-Mass Eccentric Black Hole Binaries}
\author{Katelyn J. Wagner}
\affiliation{\RIT}
\author{R. O'Shaughnessy}
\affiliation{\RIT}
\begin{abstract}
  Recent studies have shown that orbital eccentricity may indicate dynamical
  assembly as a formation mechanism for binary black holes. Eccentricity leaves
  a distinct signature in gravitational wave signals and it may be measured if the
  binary remains eccentric when it enters the LIGO band. Although eccentricity has
  not yet been confidently detected, the possibility of detecting eccentric binaries
  is becoming more likely with the improved sensitivity of
  gravitational wave detectors such as LIGO, Virgo, and KAGRA. It is
  crucial to assess the accuracy of current search pipelines in recovering
  eccentricity from gravitational wave signals if it is present. In this study,
  we investigate
  the ability of parameter estimation pipeline RIFT to recover eccentricity in
  the non-spinning and aligned-spin cases for low mass binary black holes.
  We use  \tfe{} and \teo{} to inject sets
  of synthetic signals and test how well RIFT accurately recovers key binary
  black hole parameters. Our findings provide valuable insights into the
  capability of current parameter estimation methods to detect and measure
  eccentricity in gravitational wave signals.
\end{abstract}
\maketitle
 
\section{Introduction}

Advanced LIGO \cite{2015CQGra..32g4001T} and  Virgo \cite{virgo}
ground-based gravitational wave (GW) detectors have now discovered more than
90 compact binary mergers, including a few neutron star-black hole mergers and
binary neutron star mergers \cite{gwtc3-update}. Most isolated compact
binaries that enter the LIGO band at 10Hz are expected to display
quasi-circular orbits due to gravitational radiation decaying their orbits over
time \cite{Peters_1964}. However, recent studies have suggested that some binary black
hole systems may instead display eccentric orbits before merging
\cite{Zevin_2017, Rodriguez_2018b, Samsing_2018}. These eccentric binaries are
expected to be short-lived, but they offer a unique opportunity to probe the
formation channels of binary black holes.

Viable formation channels for binary black holes fall broadly into two
categories: isolated binary evolution and dynamical assembly. The intrinsic
parameters of a binary, including component masses, component spins, and
eccentricity, can provide information about its history. Gravitational waves
can be used to probe these parameters.

The most likely formation channel for binary black holes is thought to be
through the evolution of massive binary stars---but  eccentric orbits are not
expected to result from this channel \cite{Mandel_2022}. One proposed mechanism
for forming eccentric binary black holes involves the dynamical interactions of
single black holes in densely populated stellar environments \cite{Mandel_2010},
such as  globular clusters, nuclear star clusters, or galactic
centers \cite{Rodriguez_2018a, Samsing_2014, Zevin_2019, Zevin_2021}.
In these environments, the
progenitors do not encouter each other until they have already evolved into
black holes. Then, the dynamical assembly of binaries occurs as a series of
gravitational interactions with other black holes. Interactions in such
environments are frequent, and objects may enter and leave any number of
binaries before eventually merging. The orbit is hardened in each successive
interaction and may merge while eccentricity is still present
\cite{Rodriguez_2018a}.

Understanding the formation channels of binary black holes is crucial for
interpreting gravitational wave observations and constraining theoretical
models, as different formation mechanisms leave signatures in gravitational
wave signals. The presence of eccentricity in a signal may be an excellent probe
for the dynamical formation channel \cite{Favata_2022, Lower_2018, Saini_2024},
and eccentric mergers may occur frequently
enough that gravitational waves can be used to find them.
However, detecting eccentricity in binary black hole systems is challenging due to the
low expected signal-to-noise ratio (SNR) of eccentric gravitational wave
detections. In addition, eccentric waveforms are more complicated waveforms
than those of quasi-circular systems.

Eccentricity creates distinctive and observationally-accessible modulations in
gravitational wave signals, making it key to understanding the formation and
evolution of compact binaries. As the sensitivity of ground-based detectors has increased for the next observing run, a better understanding of our ability to
recover eccentricity is vital. As for any GW detection, parameter estimation and
model selection are key in performing an analysis \cite{LIGO-CBC-S6-PE}, so an
understanding of the performance of these tools is highly important.

Several studies have attempted to infer the eccentricity of binary black hole
systems from the observed gravitational wave signals in O1 \& O2
\cite{Romero_Shaw_2019, Abbott_2019, Yun_2020, Wu_2020, Bonino_2023, OShea_2023}, in particular for GW190521 \cite{RomeroShaw_2020, Gayathri_2022, Gamba_2022},
as well as in O3 \cite{Romero_Shaw_2020, Lenon_2020, RomeroShaw_2021, Romero_Shaw_2022, Iglesias_2023, Abbott_2023} .
Many used parameter estimation and some currently available models that consider
eccentric effects: \texttt{TEOBResumS} \cite{teobresums_general}, and
\texttt{SEOBNRE} \cite{seobnre}. Another model including eccentricity is
\tfe{} \cite{taylorf2ecc}. Others have focused on high mass regimes where the
signal is much shorter so the analysis is more computationally efficient
\cite{Iglesias_2023}. However, Favata \textit{et al.} \cite{Favata_2022} show
that low-mass systems enable much sharper constraints on eccentricity and
the authors of \cite{Divyajyoti_2023} show that eccentricity is important
for statements about population characteristics and source parameter recovery.
In general, full parameter estimation may be limited
in the low mass regime as signals are long and signal-to-noise ratio is lower,
but the highly-parallelizable nature of RIFT \cite{rift, Wofford_2022} makes it an
excellent tool for this problem. Therefore, we can use RIFT to perform parameter
estimation systematics for eccentric black hole binaries using multiple waveform
models focusing on the low-mass regime to best constrain measurements of
eccentricity.

The study of eccentric binary black holes remains an active area of research,
and future detections by advanced gravitational wave detectors will provide new
opportunities to study the dynamics and formation channels of these systems.
In the meantime, we must quantitatively understand our ability to accurately
recover eccentricity as we prepare for real detections.

In this paper, we
perform a systematics study using RIFT \cite{rift, Wofford_2022} to investigate
how well eccentricity can be recovered for low-mass systems. \Secref{s:methods}
provides an overview of how RIFT performs parameter estimation. It also
describes the waveform models used in this study and details the synthetic
population created for this purpose. \Secref{s:paramDegen} discusses
degeneracies between certain parameters relevant to this study, and includes
confirmation of others' previous results using our methods. We describe the
addition of a new parameter to our pipeline, mean periapsis anomaly, which was
used in this context by \cite{Clarke_2022}. Finally, the results of our
systematics study are given in \secref{s:results}. We show recovery of
eccentric non-spinning and aligned-spin injected signals for low-mass BBH
using RIFT \cite{rift, Wofford_2022}.



\section{Methods}
\label{s:methods}

\subsection{RIFT}
\label{ss:rift}

A merging compact binary can be characterized by its intrinsic and extrinsic
parameters. The intrinsic parameters, $\lambda$, refer to the component masses,
component spins, eccentricity, and matter quantities. The seven extrinsic
paramters
($\theta$) describe the spacetime location and orientation of the system,
including right ascension, declination, luminosity distance, coalescence time,
inclination, orbital phase, and polarization.

RIFT is an iterative process consisting of two stages to estimate the intrinsic
and extrinsic parameters of the binary source. It compares gravitational wave
data to predicted signals $h_k(\pmb{\lambda},\pmb{\theta})$ where $k$ is the
number of GW detectors. In the first stage, RIFT utilizes parallel computing to
evaluate the marginal likelihood of each intrinsic parameter in a grid of
candidate points via Monte Carlo
$$
\Lmarg = \int \mathcal{L}(\pmb{\lambda},\theta)p(\theta)d\theta.
$$
This stage provides point estimates for $\ln \Lmarg$ using either Gaussian
Process (GP) regression or random forests to interpolate a full posterior
distribution over the intrinsic parameters.

In the second iterative stage, RIFT approximates $\mathcal{L}(\lambda)$ based on
the set of evaluations $\{(\lambda_\alpha, \mathcal{L}_\alpha)\}$. The data is
fitted for a full posterior distribution over the intrinsic parameters
$$
p_{\rm{post}}(\pmb{\lambda}) = \frac{\Lmarg p(\pmb{\lambda})}{\int d\pmb{\lambda}\Lmarg p(\pmb{\lambda})}
$$
where $p(\pmb{\lambda})$ is the prior on the intrinsic parameters.

The posterior is fairly sampled to generate a new grid using adaptive Monte
Carlo techniques. The grid is then fed back into the initial evaluation stage.
After multiple iterations, the log-likelihood will converge to the true value
for each parameter and the intrinsic samples will represent the true posterior
distribution. Therefore, one learns the set of true parameters of the compact
binary system.

Though RIFT is equipped to handle the non-spinning, aligned spin, and precessing
spin cases, we do not explore precession in this study due to limited 
availability of waveforms that incorporate both eccentricity and precession.

In the course of this work, RIFT was extended to include the argument of
periapsis, as this parameter is necessary to fully characterize generic
eccentric orbits. The argument
of periapsis is an angle that describes the initial orientation of a binary's
orbital ellipse which affects the gravitational wave phase. Previous
analyses using RIFT to infer the properties of black hole binaries using
semianalytic waveform models incorporating eccentricity have heretofore
adopted a fixed value for the argument of peripasis \cite{Iglesias_2023}. More
detail about mean periapsis anomoly is provided in \secref{ss:periapsis}.

\subsection{Waveform Model}
\label{ss:waveform}

In this study, we use \tfe{} and \teo{} to characterize the gravitational
radiation from black hole binaries
in eccentric orbits. We use a given model both for generating synthetic
signals and in recovering their properties with parameter inference. 

\tfe{} \cite{taylorf2ecc} is a quasi-Keplerian formalism describing post-Newtonian
(PN) corrections to binary orbits, specifically in the low eccentricity regime.
This formalism provides analytic solutions to the PN equations of motion,
computing orbital quantities $(r, \phi)$ and their derivatives as a function of
the eccentric anomoly \cite{Damour_1988, Schafer_1993}. The quasi-Keplerian
formalism enables one to track orbital evolution without computing ODEs. The
\tfe{} waveforms are inspiral only. They extend the standard circular post-Newtonian
approximations to incorporate eccentric effects for up to 3PN. The
analytic solutions are given in \cite{taylorf2ecc}. Additionally, \tfe{} is an
extension of the \texttt{TaylorF2} waveform, which is a frequency-domain
approximant evaluated in the stationary-phase approximation and assumes circular
orbits. In the $e_0 \rightarrow 0$ limit, where $e_0$ is the binary eccentricity
at reference $f_0 = 10$Hz, the \tfe{} waveform reduces to \texttt{TaylorF2}. The
value of $f_0$ is the low-frequency limit of the Advanced LIGO observing band.
This waveform model is implemented in \texttt{lalsuite} \cite{lalsuite}.

\teo{} \cite{teobresums2} is an effective-one-body (EOB) model. This approach
combines the phases of two-body dynamics, including inspiral, merger, and
ringdown, into a single analytical method. First introduced by \cite{eob},
this method allows for highly accurate waveform calculations by mapping the
dynamics of a binary system onto one single, effective object that moves in the
potential. The effective object is described by equations of motion derived from
general relativity. \teo{} is informed by NR simulations of BBH coalescence
events for calibration and validation, particularly for the eccentric version of
this model. Here, we use the eccentric spin-aligned waveforms from \teo{}
\cite{teogeneral_update} which were verified by NR for eccentricities up to
$e \leq 0.3$ at $10$ Hz. This model also incorporates higher order modes up
to $l = |m| = 5$, except for $m=0$. The user can also define the reference
frequency for \teo{} at periastron, apastron, or the mean of both. 

As mentioned above, time-dependent quantities like eccentricity are typically
defined
in terms of a reference frequency. Most eccentric models then use a reference
eccentricity, $e_0$, that has a particular defined value at some reference
frequency $f_{\rm{ref}}$.
%
For \tfe{},  the input eccentricity and reference frequency are used to
compute additional relative PN correction terms to third order. For \teo{}, $e_0$
and $f_{\rm{ref}}$ are used to determine the initial conditions that are used to
evolve the trajectories of the binary components. The value of the waveform
eccentricity is different than the physical, time-varying eccentricity of the
system, and may also mean different things depending on the choice of model.

In this proof-of-concept study, we do not attempt to carefully standardize the
definition of eccentricity between the different
models used as in \cite{Shaikh_2023}; we merely demonstrate how well the
eccentricity parameter adopted in each model can be constrained by our parameter
estimation pipeline.

The waveforms described above only incorporate leading-order quadrupole
radiation. During periapsis, highly eccentric compact binaries close to merger
should also emit strongly in nonquadrupole modes. Some proof-of-concept studies
have demonstrated how to infer parameters of generic eccentric sources including
these higher-order modes, particularly in the context of numerical relativity
simulations. The sources of interest in this work are low mass,
long-duration sources  whose strongest nonquadrupole radiation is largely at
frequencies at-or-above the peak sensitivity of our detector network. As
such we expect higher-order modes are relatively less important. As was done in
\cite{Iglesias_2023}, for \teo{} we include ${\ell}_{\rm{max}} \leq 4 $ except
the $m = 0$ modes.


\subsection{Synthetic Population and PP tests}
\label{ss:synthpop}

In this work, we use RIFT to infer the parameters of sets of populations of
synthetic sources. The ranges and priors for key signal parameters are provided
in \tabref{t:injections}. Values for each of a signal's parameters are drawn
randomly from the specified range and according to the defined prior.

We focus on recovering eccentricity in the low mass regime for binary black
holes. The relevant parameter ranges for our synthetic population are listed in
\tabref{t:injections}. The individual source masses are drawn uniformly in
$m_i$ in a region bounded by $\mc / M_\odot \in [10,20]$ and
$\eta \in [0.2, 0.25]$. The sources are placed at a luminosity distance of
$500-1000$ Mpc, drawn proportionally to $d_L^2$. The eccentricity of each
synthetic binary falls within $e_0 \in [0.01, 0.1]$.

For eccentric sources, we first investigate the non-spinning case and then
consider aligned-spin sources, with each spin component assumed to be uniform in
$[-0.5,0.5]$. We sample over aligned-spin using an effective spin parameter
$$
\chi_{\rm{eff}} = \frac{(m_1 \chi_{1,z} + m_2 \chi_{2,z})}{(m_1+m_2)}.
$$
Spin precession is
not considered in this study. This is primarily due to a lack of available
waveforms including both eccentricity and precession, as discussed in the
previous section. 


\begin{table}[]
  \centering
  \label{t:injections}
  \begin{tabular}{|c||c|c|c|}
    \hline
    Parameter & Symbol & Prior & Injected Range \\
    \hhline{|=|=|=|=|}
   Chirp Mass   &  $\mc$  & uniform in $m_i$ &  $[10-20] M_\odot$  \\ \hline
    Distance  &  $d$ & $\propto d_L^2$  & $[500-1000]$ Mpc  \\ \hline
    Eccentricity &  $e_0$ & uniform & $[0.01,0.1]$  \\ \hline
    Spin  &  $\chi_{\rm{eff}}$ & uniform & $[-0.5,0.5]$  \\ \hline
    Inclination & $i$ & uniform & $[0.0, \pi]$ \\ \hline
  \end{tabular}
  \caption{Ranges and priors for key parameters of synthetic signal set. The values for $\chi_{\rm{eff}}$ apply only for the aligned spin case, as $\chi_{\rm{eff}}$ is set to zero for the non-spinning tests.}
\end{table}

For reproducibility, we simulate signals for eccentric BBH in a three-detector
network with $4096$Hz time series in Gaussian noise with known aLIGO design
PSDs. The same noise realization is used for all analysis runs. We begin the
signal evolution and the likelihood integration at $20$Hz. Each source has
an SNR of $\sim 10-20$.

Probability-probability (PP) plots are one way to evaluate the performance of
parameter inference \cite{pp_plots}. Using RIFT on each source $k$ with true
signal parameters $\lambda_k$, we estimate the fraction of the posterior
distributions which is below the true source value $\lambda_{k,\alpha}[\hat{P}_{k,\alpha}(< \lambda_{k,\alpha})]$
for each intrinsic parameter $\alpha$. After reindexing the sources such that
$\hat{P}_{k,\alpha}(\lambda_{k,\alpha})$ increases with $k$ for some fixed
$\alpha$, a plot of $k/N$ versus $\hat{P}_{k}(\lambda_{k,\alpha})$ for relevant
binary parameters can be compared with the expected result $(P(<p)=p)$ and the
binary uncertainty interval. Such PP plots will be shown in the results in \secref{s:results}.


\section{Parameter Degeneracies}
\label{s:paramDegen}

\subsection{Chirp Mass}
\label{ss:chirpMass}

The authors of \cite{Favata_2022} explore a degeneracy between eccentricity
and chirp mass for \tfe{} and introduce a new parameter called
\textit{eccentric chirp mass}. This parameter explains a bias seen in the
standard chirp mass parameter when eccentricity is present in the waveform. As
binary eccentricity $e_0$ increases, the bias in recovered chirp mass
$\mc$ grows, tending to recover a higher chirp mass than the
injected value. See Figures 3 \& 4 in reference \cite{Favata_2022}.

The standard chirp mass for circular binaries is
\begin{equation}
  \mc \equiv \eta^{3/5} M_{\rm{tot}} = (m_1 m_2)^{3/5} (m_1+m_2)^{-1/5}
\end{equation}
where $\eta$ is the symmetric mass ratio. However, \cite{Favata_2022} defines
eccentric chirp mass for low-eccentricity binaries as
\begin{equation}
  \label{e:mcecc}
  \mcecc = \frac{\mc}{(1 - \frac{157}{24}e_0^2)^{3/5}}
\end{equation}
which serves to replace the standard ``circular'' chirp mass and explain a
degeneracy between eccentricity and standard $\mc$. We find
similar results using RIFT.

\begin{figure}[htbp]
    \includegraphics[width=0.45\textwidth]{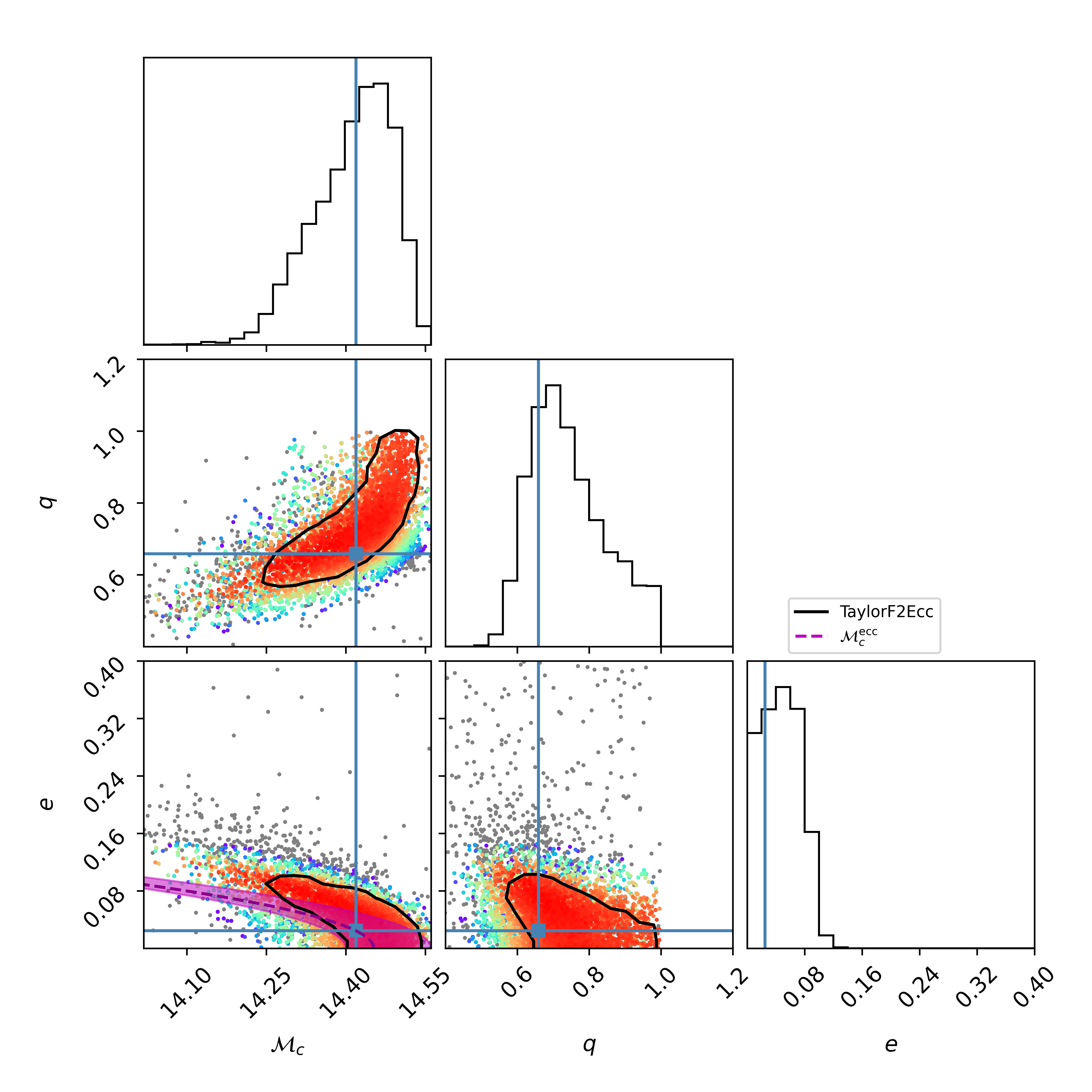}
    \caption{A corner plot showing injection results for one synthetic signal event. The true event parameters are shown by solid blue lines. To test the
      eccentric chirp mass parameter, we plot the value of
      $\mcecc$ calculated using \eqnref{e:mcecc} and the true values of
      $\mc$ and $e_0$ on the 2D recovered eccentricity-chirp mass posterior.
      We see that the eccentric chirp mass parameter better fits the peak of
      the recovered posteriors,
      demonstrating the expected behavior as described by \cite{Favata_2022}.
      Note the strong parameter correlation between eccentricity and chirp mass,
      shown by the recovered points and calculated by the true $\mcecc$ line.}
    \label{fig:mcecc}
\end{figure}

Measurements of $\mc$ are strongly correlated
with eccentricity. We can now better constrain chirp mass using the parameter
$\mcecc$ introduced in \cite{Favata_2022}.  $\mcecc$ is a natural parameter that
better corresponds to the shape of the $\mc$ posterior. For this reason, we
implement $\mcecc$ as an additional fitting coordinate in RIFT.

\figref{fig:mcecc} shows that we find the same degeneracy between eccentricity
and chirp mass as discussed by \cite{Favata_2022}. In this figure we use
show a fit for $\mcecc$ superimposed on
our standard chirp mass posteriors after RIFT has completed parameter recovery.
The points indicate individual search points in parameter spaced and are mapped
to colors based on their likelihood values. Red points indicate the highest
likelihoods. We plot the truth values for the injected signal as solid blue
lines. The peak in the recovered posterior for chirp mass falls slightly above
the injected value. When we calculate $\mcecc$ based on the true values for
$\mc$ and $e_0$, the magenta dashed line shows
that the eccentric chirp mass more accurately finds the peak of the recovered
posterior as shown in \figref{fig:mcecc}. 


\subsection{Argument of Periapsis}
\label{ss:periapsis}

The argument of periapsis at $\omega_{\rm{ref}}$. This parameter is the fraction
as an angle of the orbital period through which the system has rotated since the
last pericenter passage. This angle is measured from the reference plane of a
binary orbit to the point of periapsis passage. Eccentricity and mean anomaly
are the two additional intrinsic parameters that are required to fully describe
the orbit of an eccentric binary black hole system \cite{Shaikh_2023}.

The waveform reference frequency $f_{\rm{ref}}$ and periapsis the reference
frequency $\omega_{\rm{ref}}$ may not be the same when performing signal
recovery. Since values of eccentricity and mean anomaly vary during binary
evolution, the authors of \cite{Shaikh_2023} advocate for standardized
definitions of a reference point at which to measure the values of these
parameters, and make suggestions for how to do so.

The two waveform models used in this investigation do not allow the user to
independently vary the argument of periapsis. \cite{Clarke_2022} develops
method to examine $\omega_{\rm{ref}}$ indirectly for \teo{}. The authors conclude
that analyses to date are unlikely to have suffered from fixing a value for this
parameter given the relatively low SNR of eccentric candidate sources. They do
however recommend that future analyses marginalize over this parameter as
detectors become more sensitive to potentially eccentric sources and until more
waveform models become available. We also note the recent investigation
\cite{Ramos_2023} which performed proof-of-concept parameter inference using a
new effective one-body model which implements this parameter natively.



The argument of periapsis is now availble in RIFT to be used with future
waveform models that include it, in preparation to eventually test how this
parameter affects parameter estimation. However, the two models used in this
study do not allow direct control of this parameter so we do not use it to
measure eccentricity at this time.


\section{Results}
\label{s:results}

In this section, we use RIFT to recover the parameters of ensembles of 100
sources whose properties were described in \secref{ss:synthpop} and
which were created using the models described in \secref{ss:waveform}.

The parameter ranges for the injected signals are shown in
\tabref{t:injections}. We investigate both the non-spinning and aligned-spin
cases for eccentric sources to demonstrate RIFT's ability to recover
eccentricity well.

\begin{figure}[bhtp!]
    \begin{minipage}[t]{0.9\columnwidth}
        \includegraphics[width=\textwidth]{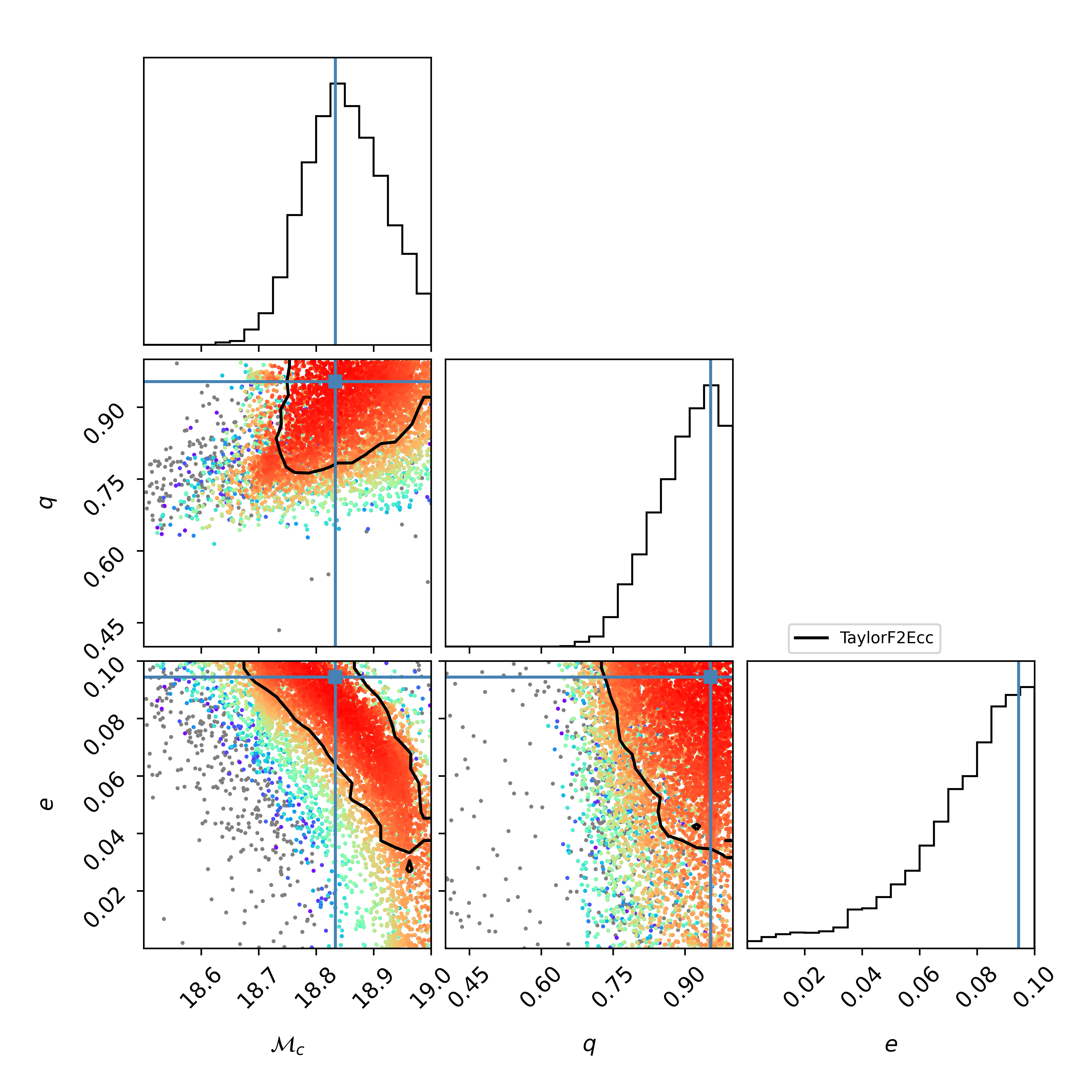}
        \label{fig:tfe_nospin}
    \end{minipage}
    \hfill
    \begin{minipage}[t]{0.9\columnwidth}
      \includegraphics[width=\textwidth]{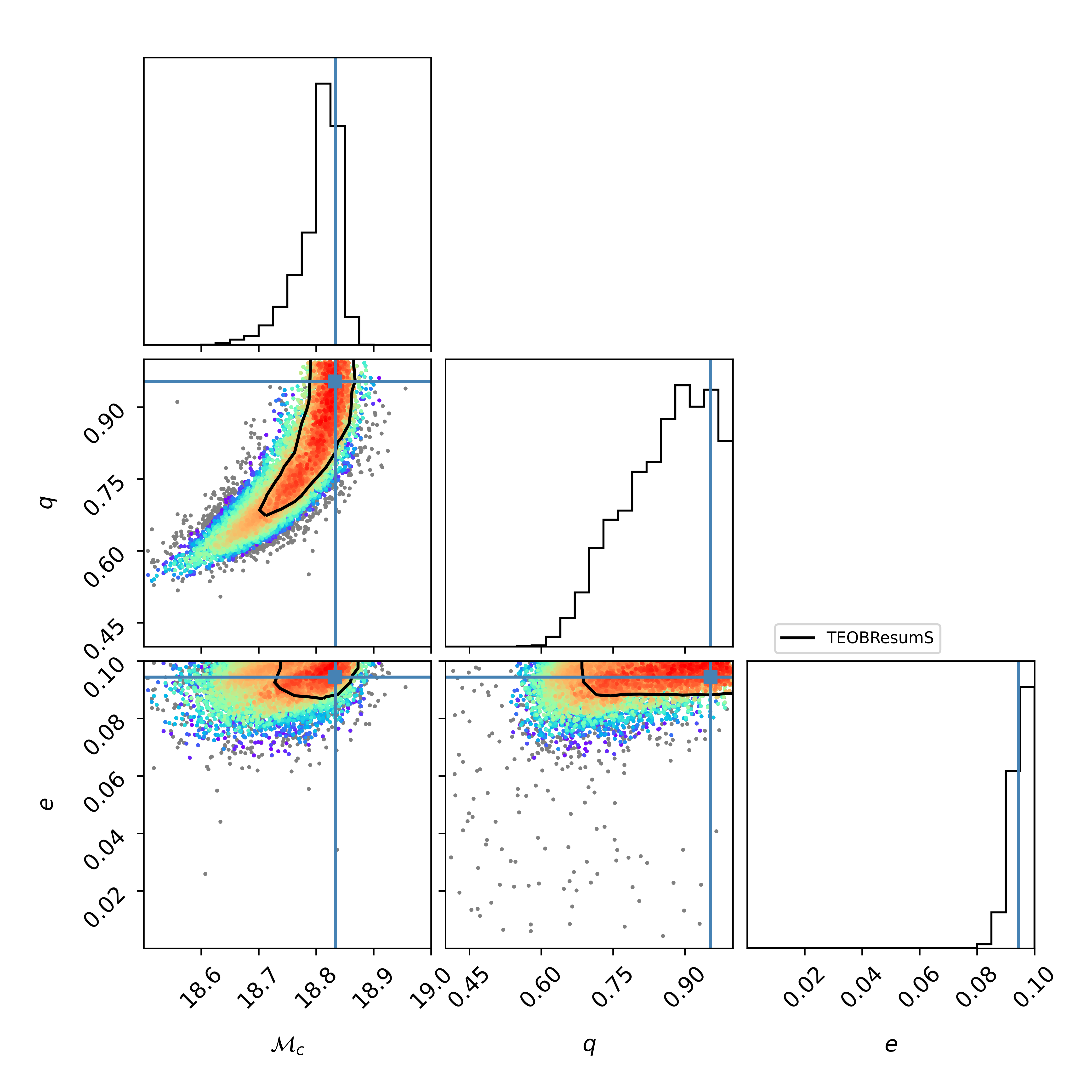}
      \label{fig:teo_nospin}
    \end{minipage}
    \caption{Top: Corner plot showing one and two dimensional marginal
      distributions $\mc, q, e$ for a signal recovered with \tfe{}. The
      injected signal has no spin. The contours represent the 90\% confidence
      intervals for each joint distribution. Solid blue lines indicate true
      parameter values. This plot show $\mc$ contours rather than $\mcecc$
      contours. Bottom: corner plot showing one and two dimensional marginal
      distributions $\mc, q, e$ for a signal recovered with \teo{}. The
      injected signal has no spin. Note that the signal parameters are
      identical for these two plots (besides injected and recovered waveform
      model).}
    \label{fig:corner_nospin}
\end{figure}

\begin{figure*}[htbp!]
\centering
\subfloat[\tfe{} injection and recovery\label{fig:tfe_diag_mc}]{
\includegraphics[width=0.45\textwidth]{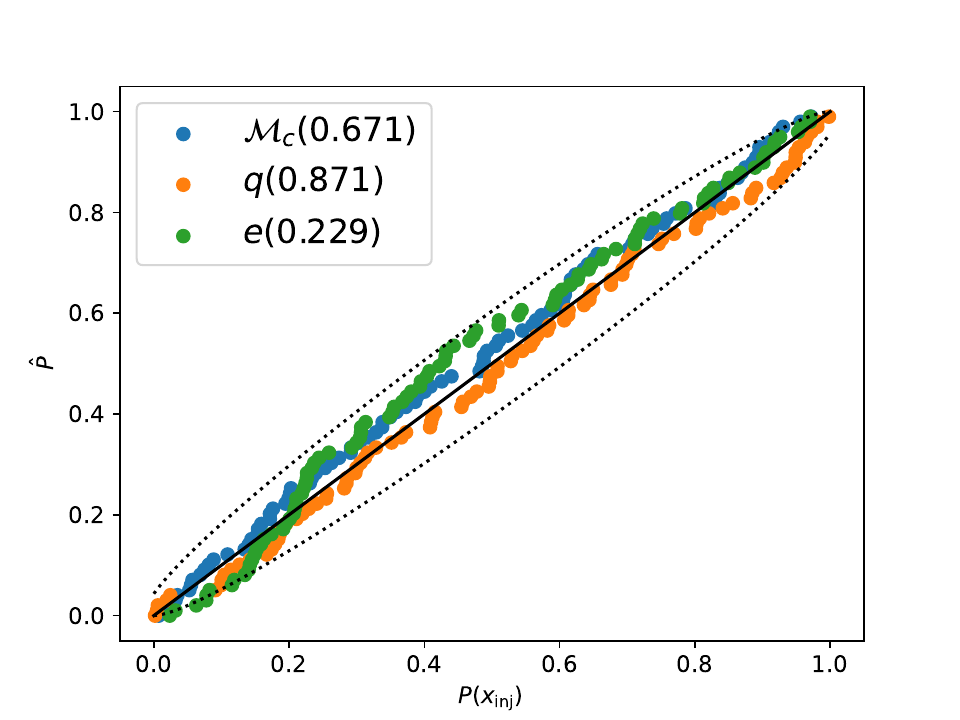}}
\quad
\subfloat[\teo{} eccentric injection and recovery \label{fig:teo_diag}]{
\includegraphics[width=0.45\textwidth]{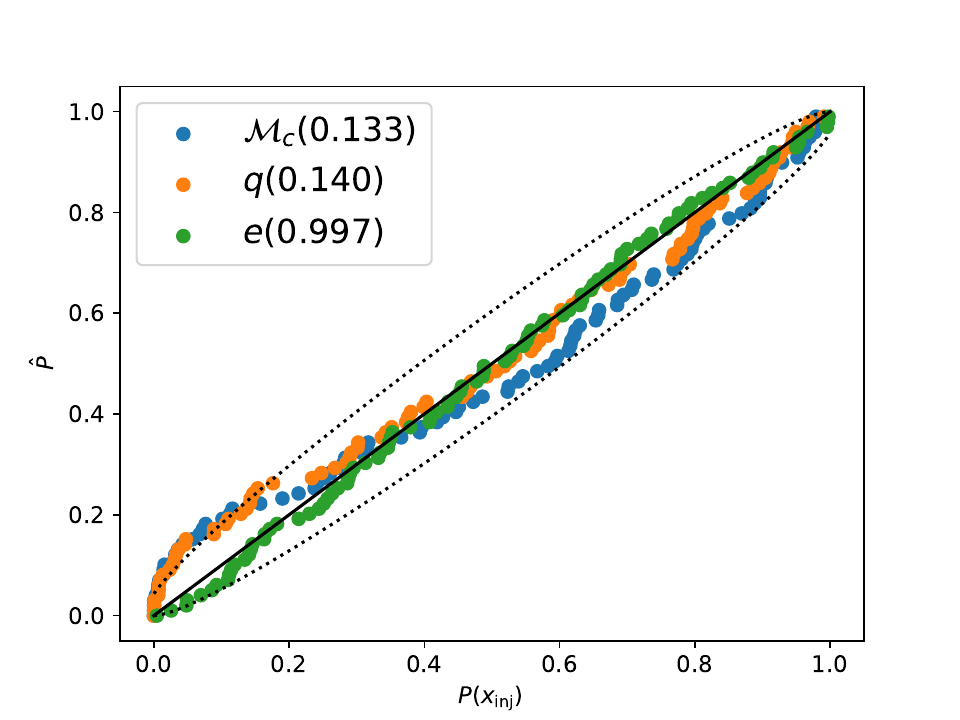}}
\quad
\subfloat[\texttt{TaylorF2} recovery of eccentric templates \label{fig:tfe_nondiag}]{
\includegraphics[width=0.45\textwidth]{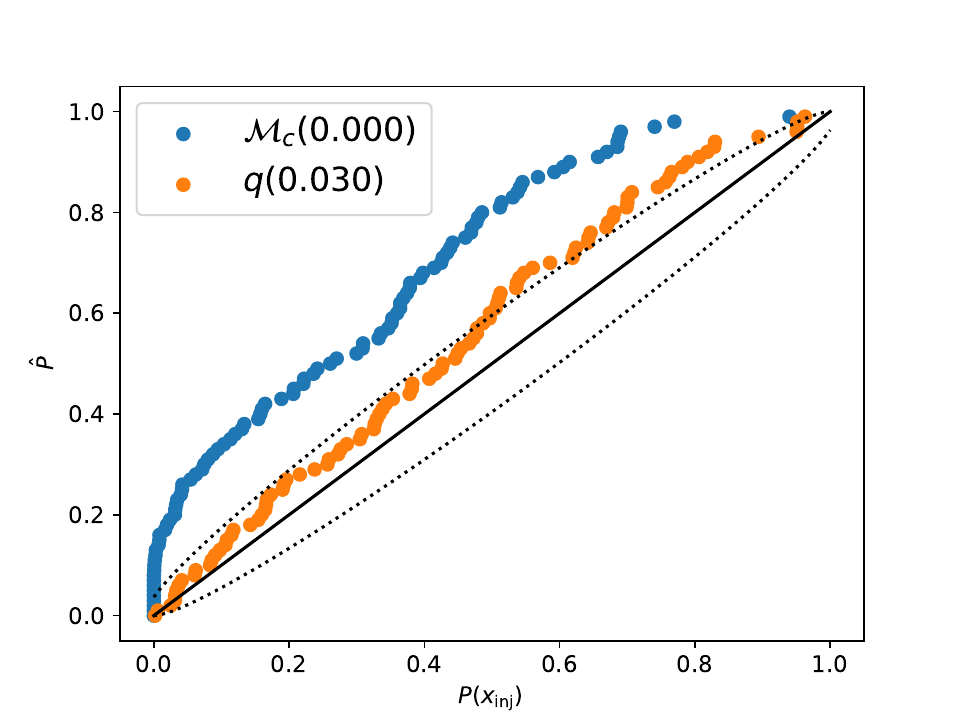}}
\quad
\subfloat[\teo{} eccentric injection, quasi-circular recovery templates \label{fig:teo_nondiag}]{
\includegraphics[width=0.45\textwidth]{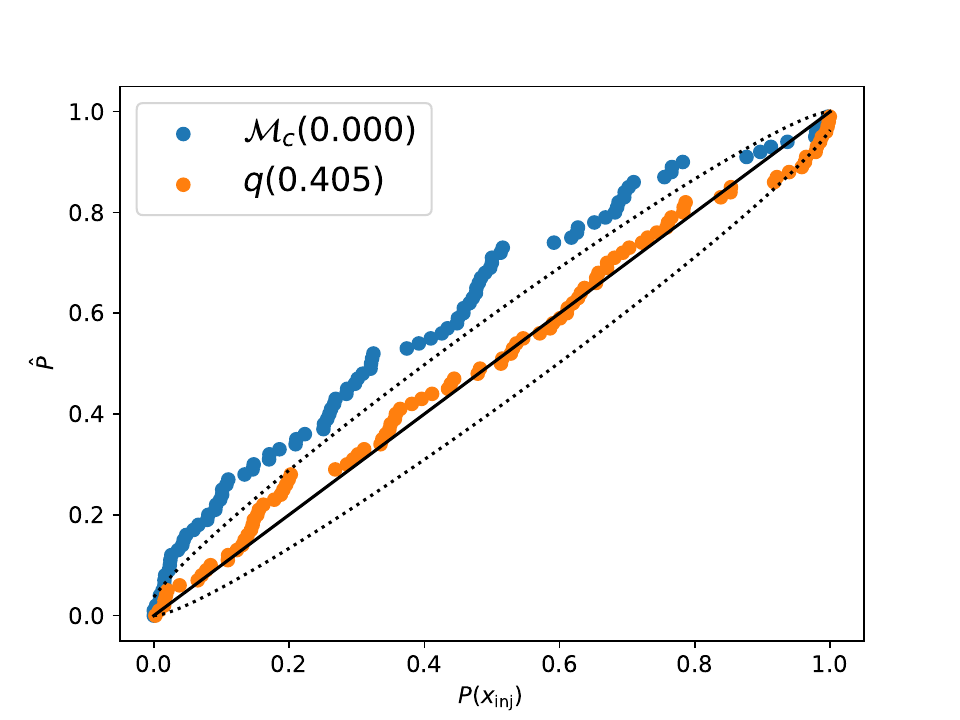}}    
\caption{Top-left: PP-plot of non-spinning eccentric events injected with \tfe{}
  and recovered with \tfe{}. Top-right: PP-plot of non-spinning eccentric events
  injected and then recovered with \teo{} including eccentricity. Bottom-left:
  Recovery of same set of non-spinning eccentric events injected with \tfe{},
  but recovered with the quasicircular waveform model \texttt{TaylorF2}, which
  is equivalent to \tfe{} for $e=0$. Bottom-right: Recovery of same set of
  non-spinning eccentric events injected with \teo{} that include eccentricity,
  but recovered with the non-eccentric version of \teo{}. For all plots, the
  dashed line indicates the 90\% credible interval expected for a cumulative
  distribution drawn from 100 uniformly-distributed samples. All panels show
  recovery of the same injection set (i.e. all injection parameters are
  identical per event except for the waveform model), demonstrating that
  eccentric physics must be included in recovering eccentric waveforms
  regardless of waveform model.}
\label{fig:ns_pp_plots}
\end{figure*}

\subsection{Eccentric Non-spinning Recovery}
\label{ss:nospin}

We first test the ability of RIFT to recover injected signals that include
eccentricity but where the binary components are non-spinning. In
\figref{fig:corner_nospin}, the top panel shows an example of \tfe{}
\cite{taylorf2ecc} used for both injection and recovery, and the bottom panel
shows an example of \teo{} \cite{teobresums2} injection and recovery of a
waveform with identical parameters as the top panel. We see that the signal
is well recovered in both cases, but note definitive differences between the
shapes of the posteriors depending on which waveform was used.

In this recovery set, we can use $\mcecc$ to fit the chirp mass posteriors, but
we do not use this parameter for sampling. Additionally, we obtained good
recovery results without the use of the argument of periapsis, which we do not
use for our recovery of non-spinning eccentric signals
because it is not available in the waveform implementations for
either \tfe{} or \teo{}.

We use a PP-plot to examine whether the signal parameters are recovered well
over the whole set of synthetic signals. A diagonal PP-plot indicates that the
distribution of the recovered signals matches the distribution of those
injected, as described in \secref{ss:synthpop}. For our non-spinning eccentric
signals, we obtain a diagonal PP-plots shown in \figref{fig:ns_pp_plots}: our
inference reliably recovers the source parameters, including eccentricity.

PP-like plots also allow us to demonstrate conclusively just how
important eccentricity is when characterizing source properties.
\figref{fig:tfe_diag_mc} shows that our parameter inference is unbiased for
\tfe{}.
Then we repeat the analysis described above, but require exactly quasicircular
orbits during inference: we recover eccentric non-spinning signals with a
non-eccentric, non-spinning waveform. Here we
use \texttt{TaylorF2}, which is the same as the \tfe{} model used above but
excludes eccentric corrections. As seen in \figref{fig:tfe_nondiag}, the
PP-plot is no longer diagonal:  eccentric physics must be included to infer any
source properties even for a population of modest-amplitude signals with
observationally accessible eccentricities.

The same process is performed again using \teo{}. For this analysis we include
higher order modes up to $\ell_{\rm{max}}=4$ except for $m=0$ according to the
available subdominant modes for this waveform. \figref{fig:teo_diag} shows that
our parameter inference is unbiased for this second waveform as well.
Then we once again repeat the analysis, but require quasicircular templates from
\teo{} during inference: we recover a  eccentric (non-spinning) signals made
with the eccentric version of \teo{} with the non-eccentric, non-spinning
version of \teo{}. As seen in \figref{fig:teo_nondiag}, the PP-plot is no longer
diagonal. This reiterates that eccentric physics must be included to infer any
source properties --- even for a population of modest-amplitude signals with
observationally accessible eccentricities for both waveform models we test.

One interesting feature to note is that the strong correlation between $\mc$ and
$e$ present in the \tfe{} waveform, as seen in \figref{fig:corner_nospin} and
described by \cite{Favata_2022} as $\mcecc$, is missing from the \teo{}
recovery, where chirp mass and eccentricity appear to be fairly uncorrelated.


\begin{figure}[htbp!]
    \begin{minipage}[t]{0.9\columnwidth}
        \includegraphics[width=\textwidth]{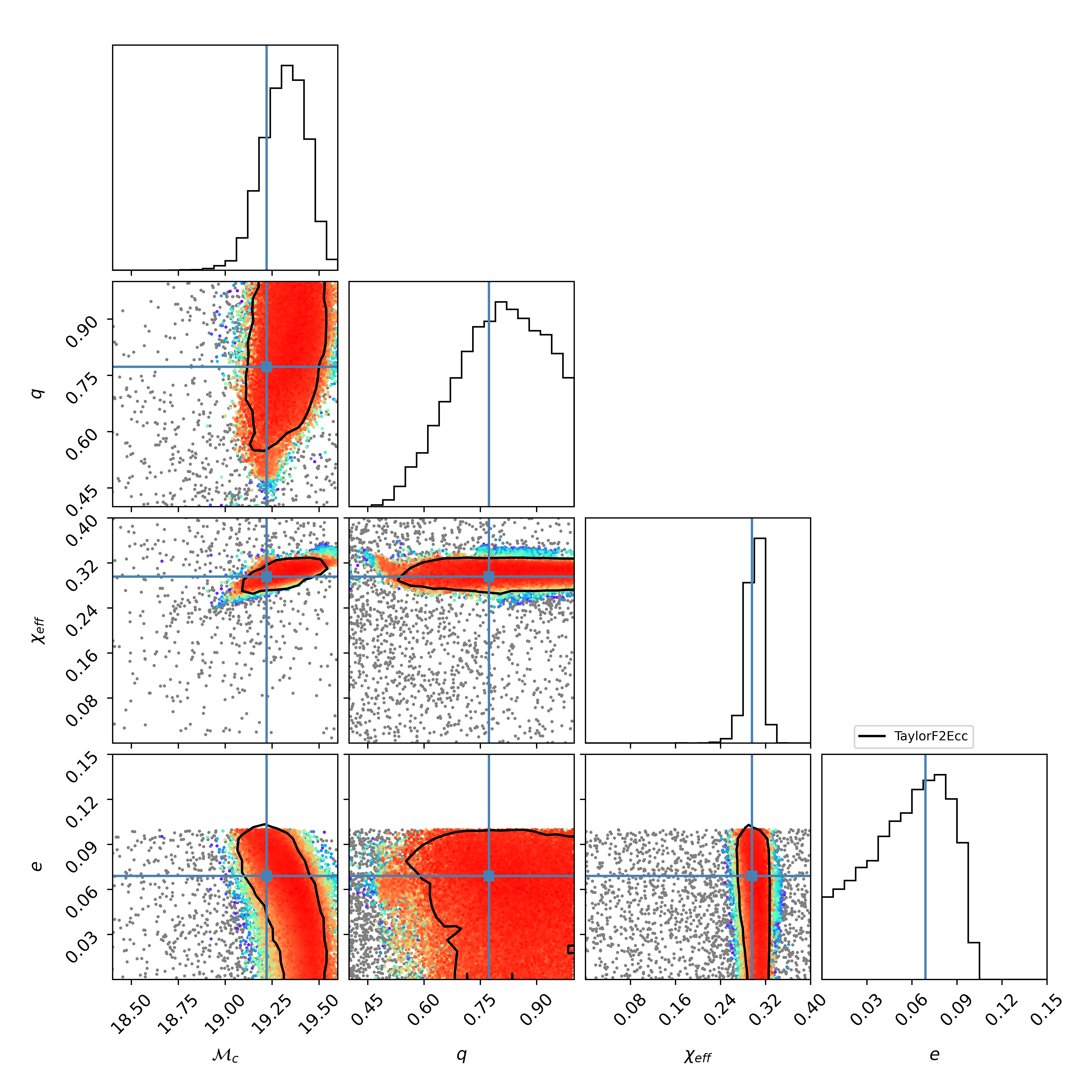}
        \label{fig:tfe_as}
    \end{minipage}
    \hfill
    \begin{minipage}[t]{0.9\columnwidth}
      \includegraphics[width=\textwidth]{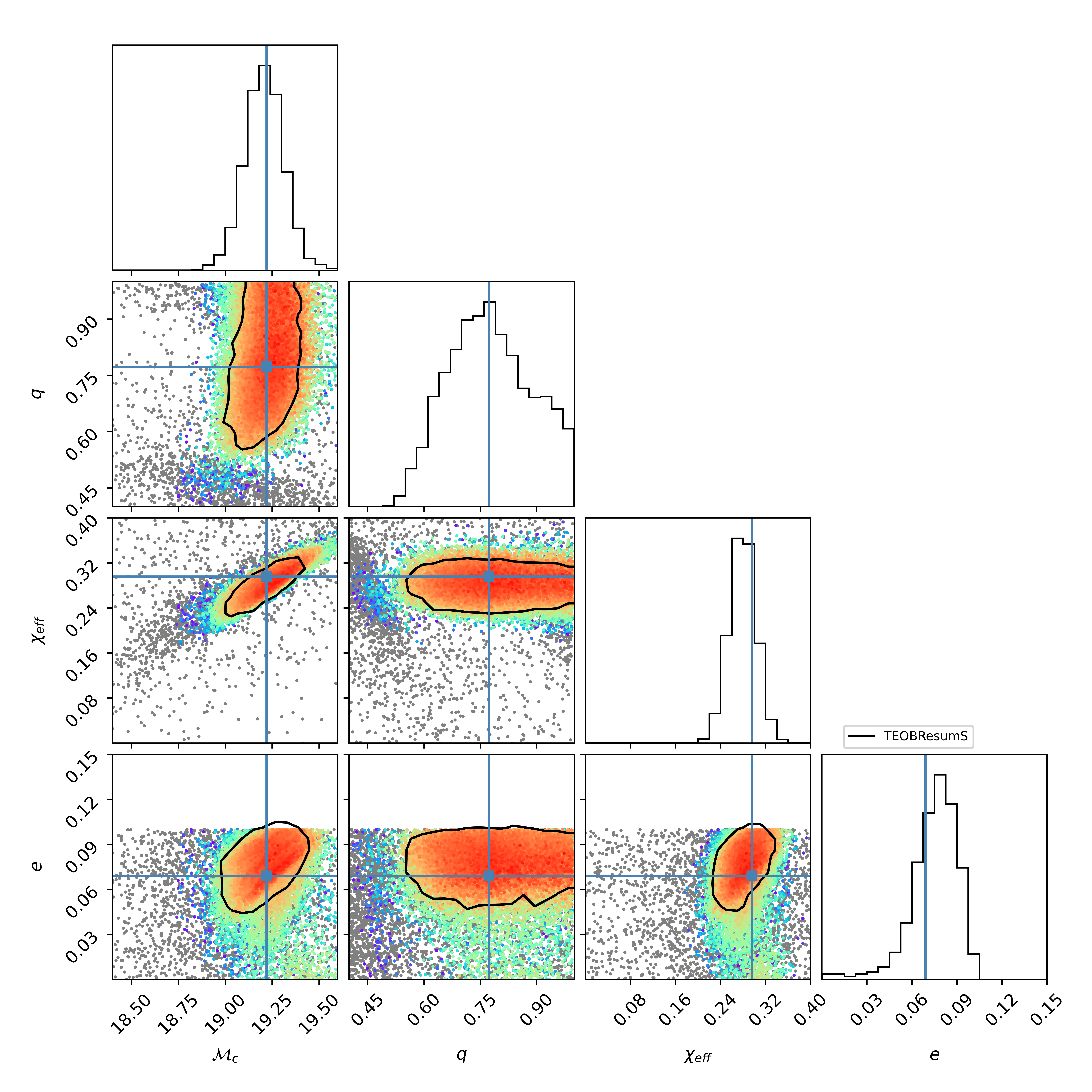}
      \label{fig:teo_as}
    \end{minipage}
    \caption{Top: Corner plot showing one and two dimensional marginal
      distributions $\mc, q, \chi_{\rm{eff}}, e$ for a signal recovered with
      \tfe{}. The injected signal has aligned spin. The contours represent
      the 90\% confidence intervals for each joint distribution. Solid blue
      lines indicate true parameter values. Bottom: corner plot showing one
      and two dimensional marginal  distributions $\mc, q, \chi_{\rm{eff}}, e$
      for a signal recovered with \teo{}. The injected signal has aligned spin.
      Note that the signal parameters are identical for these two plots
      (besides injected and recovered waveform model).}
    \label{fig:corner_as}
\end{figure}

\subsection{Eccentric Aligned-Spin Recovery}
\label{ss:aligned}

Similarly, we test the ability of RIFT to recover injected signals that include
eccentricity where the binary components have aligned spin vectors. In
\figref{fig:corner_as}, we use \tfe{} \cite{taylorf2ecc} and \teo{} to both
inject and recover these waveforms to see that the signal is well recovered.

\begin{figure}[htbp!]
    \begin{minipage}[t]{0.9\columnwidth}
        \includegraphics[width=\textwidth]{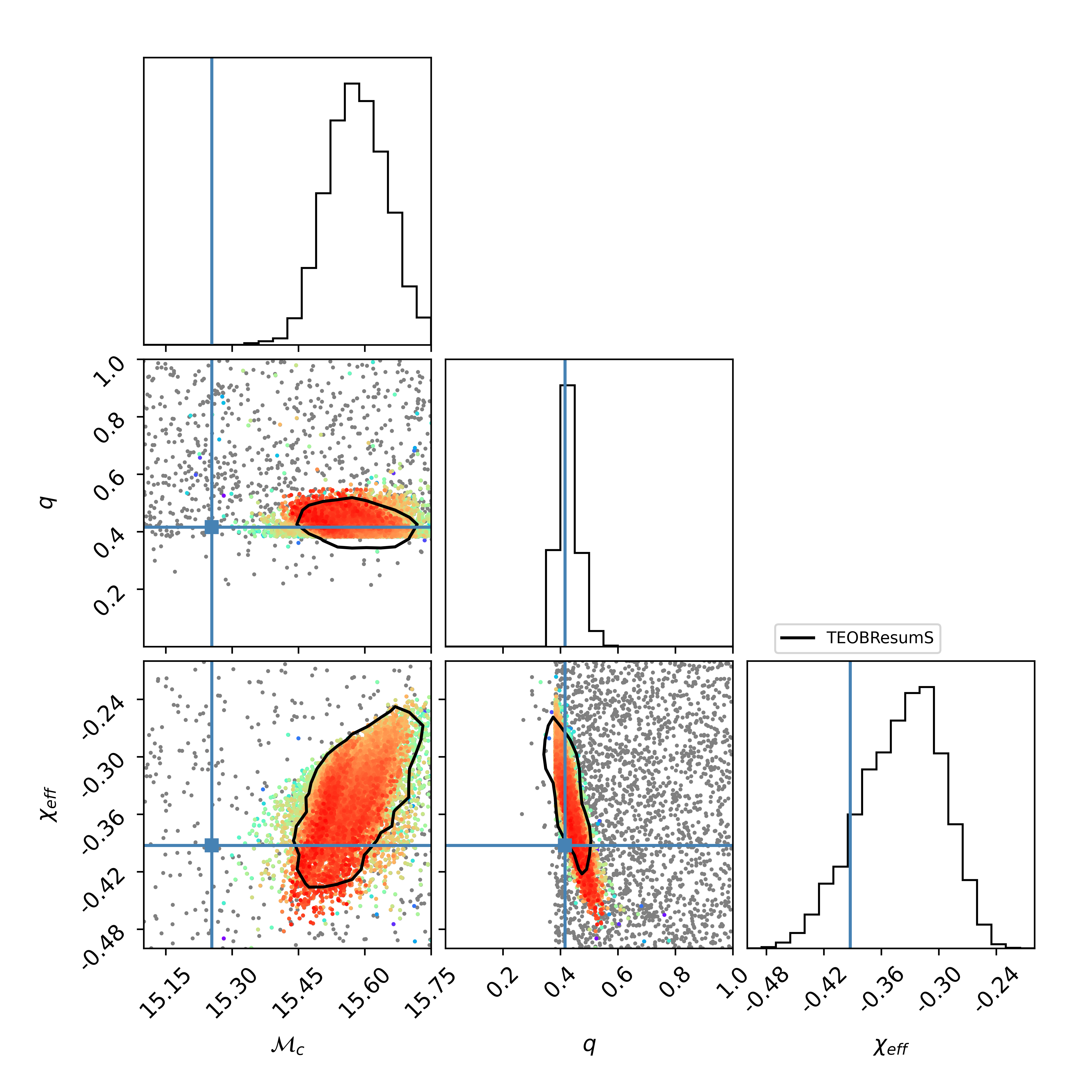}
        \label{fig:teo_noe_bias}
    \end{minipage}
    \hfill
    \begin{minipage}[t]{0.9\columnwidth}
      \includegraphics[width=\textwidth]{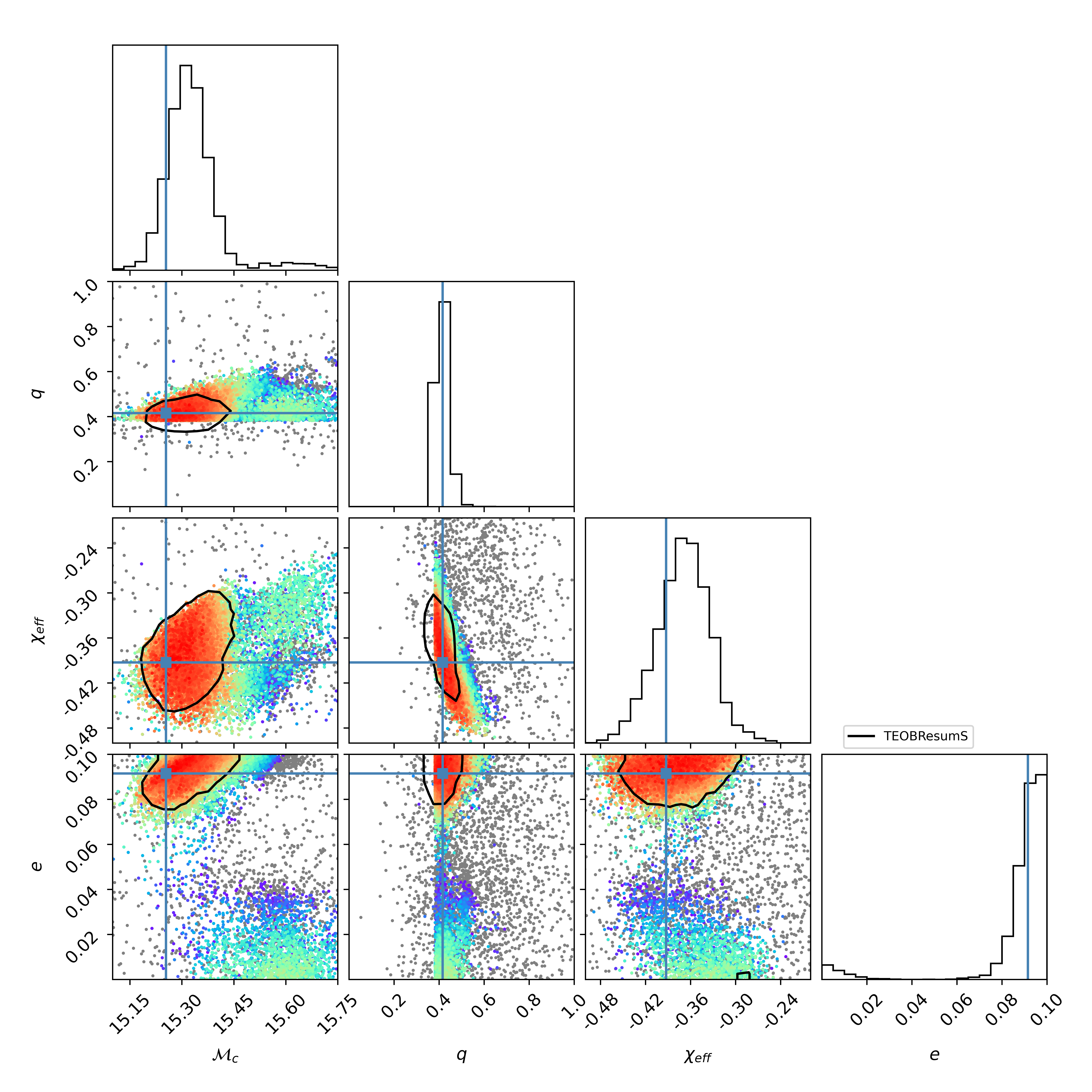}
      \label{fig:teo_ecc_bias}
    \end{minipage}
    \caption{Top: Corner plot showing one and two dimensional marginal
      distributions $\mc, q, \chi_{\rm{eff}}$ for an eccentric signal recovered
      with a quasi-circular \teo{} template. The injected signal has aligned
      spin. Note the poor recovery of $\mc$ when eccentricity is not considered
      in signal recovery. Bottom: corner plot showing one
      and two dimensional marginal  distributions $\mc, q, \chi_{\rm{eff}}, e$
      for \textit{the same} eccentric signal recovered with an \textit{eccentric} \teo{}
      template. The accuracy of the recovered signal parameters drastically
      improves when eccentric recovery methods are used.}
    \label{fig:corner_bias}
\end{figure}

As an anecdotal example of the importance of including eccentricity in signal
recovery, we show the bias that can arise in recovered parameter values in
\figref{fig:corner_bias}. We choose an example signal near the upper end of our
range in eccentricity. The top corner plot shows an eccentric \teo signal
recovered with quasi-circular waveform template. In this example, we attempt to
recover the parameter values of an eccentric signal without considering
eccentricity. We see that the recovery completely misses the value of the chirp
mass. However, when we do consider eccentricity in the recovery of this same
eccentric signal, as shown in the bottom corner plot of \figref{fig:corner_bias},
the values of the parameters are much more accurately recovers. This illustrates
the importance of the inclusion of eccentricity in the recovery of eccentric
signals.

\begin{figure*}[htbp!]
\centering
\subfloat[\tfe{} injection and recovery\label{fig:tfe_diag_as}]{
\includegraphics[width=0.45\textwidth]{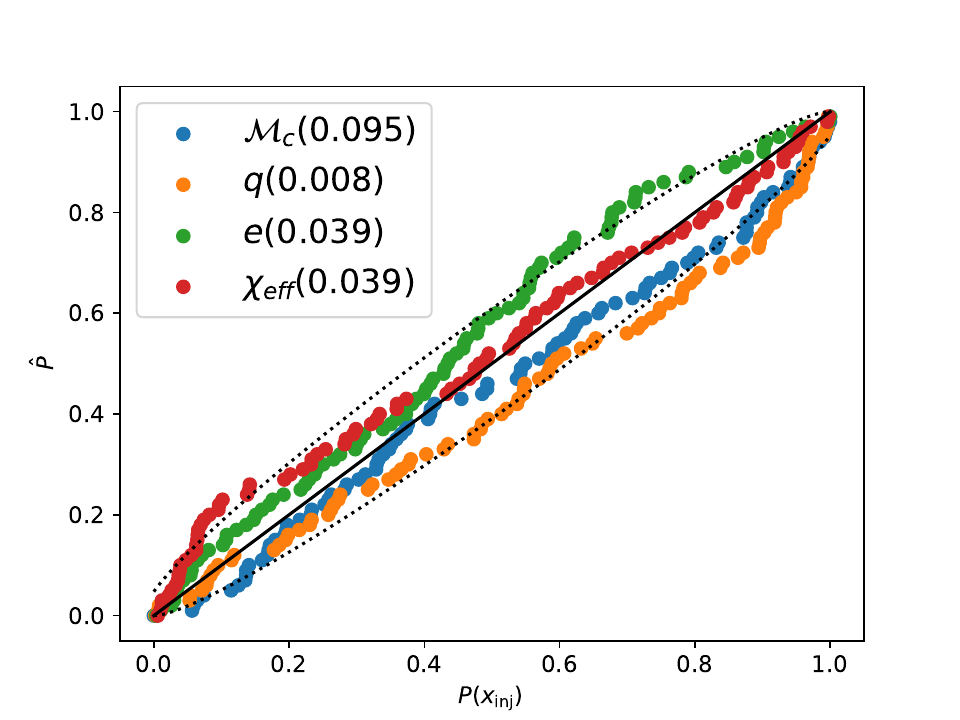}}
\quad
\subfloat[\teo{} eccentric injection and recovery \label{fig:teo_diag_as}]{
\includegraphics[width=0.45\textwidth]{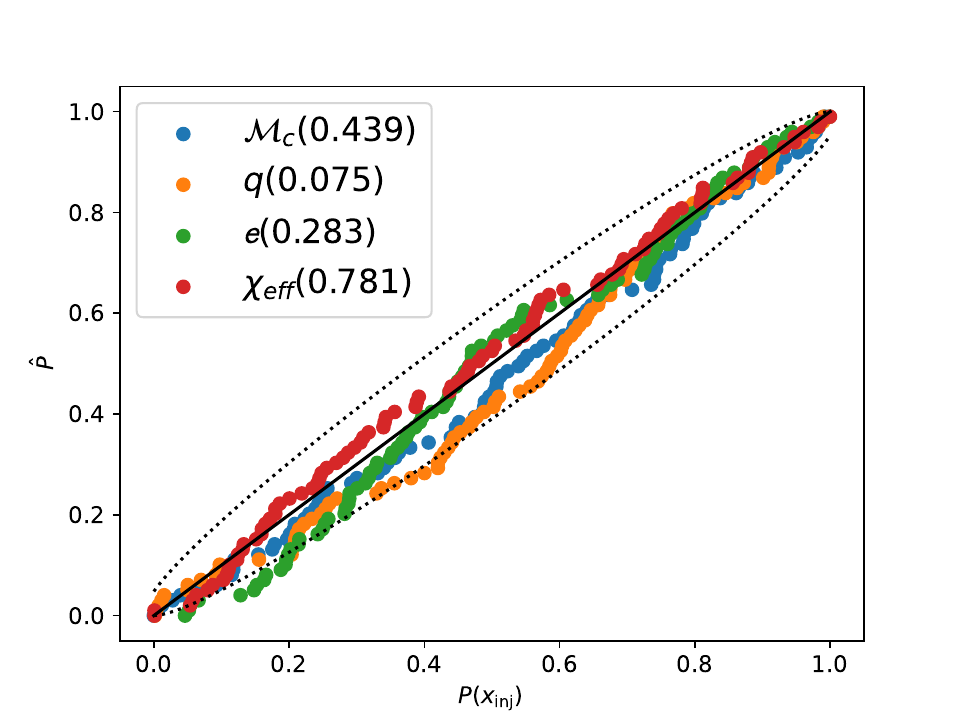}}
\quad
\subfloat[\texttt{TaylorF2} recovery of \tfe{} templates \label{fig:tfe_nondiag_as}]{
\includegraphics[width=0.45\textwidth]{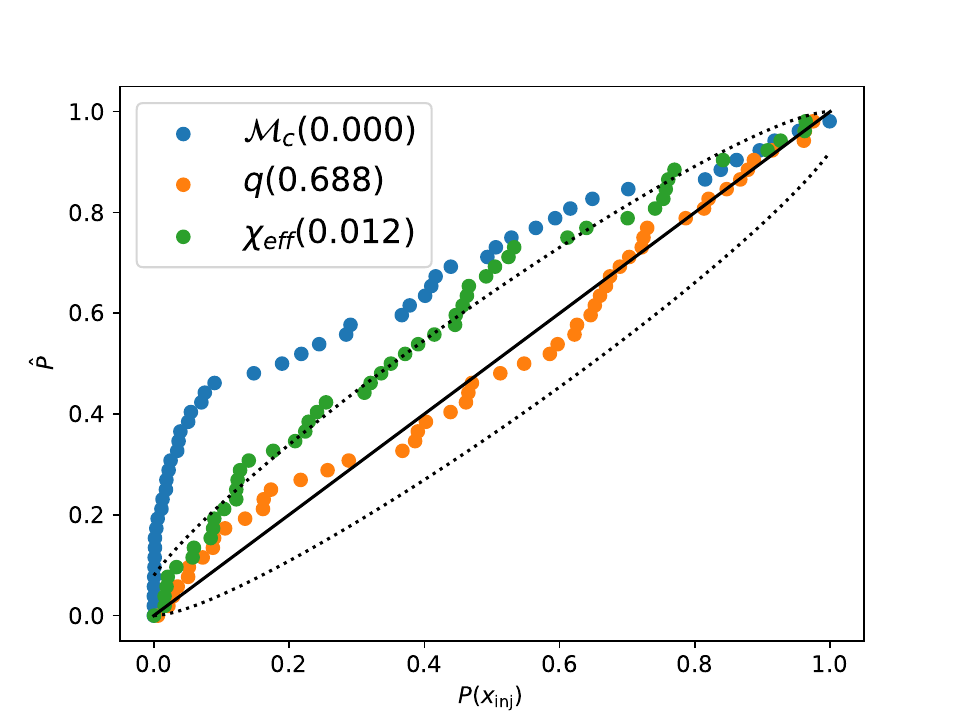}}
\quad
\subfloat[\teo{} quasi-circular recovery of eccentric injections \label{fig:teo_nondiag_as}]{
\includegraphics[width=0.45\textwidth]{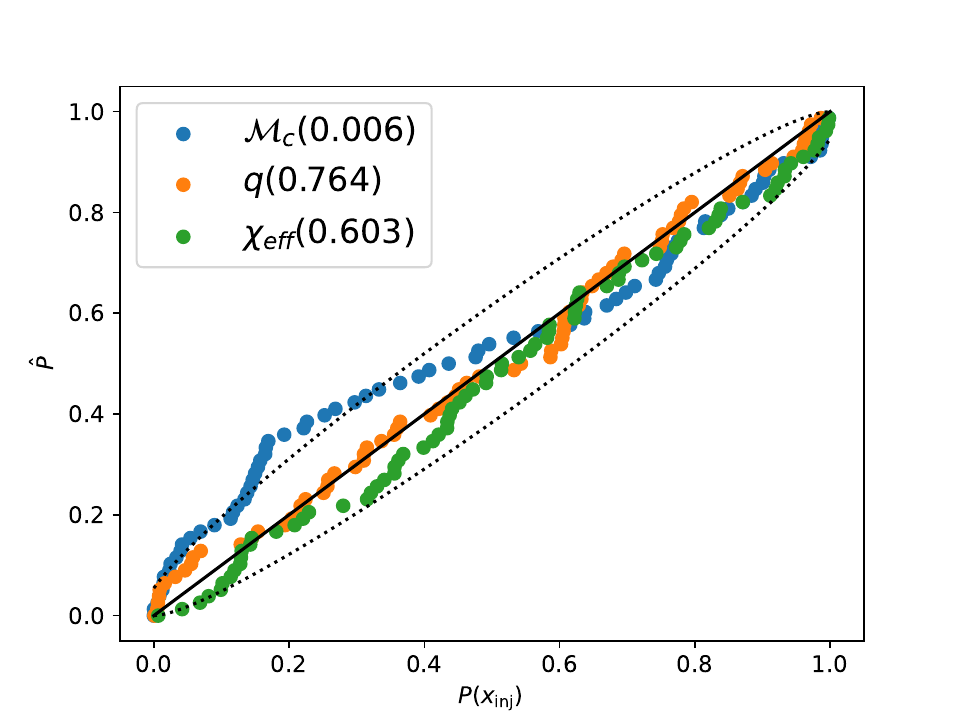}}    
\caption{Top-left: PP-plot of aligned-spin eccentric events injected with \tfe{} and recovered with \tfe{}.
  Top-right: PP-plot of aligned-spin eccentric events injected and then recovered with \teo{} including
  eccentricity. Bottom-left: Recovery of same set of aligned-spin eccentric events injected with \tfe{},
  but recovered with the quasicircular waveform model \texttt{TaylorF2}, which is equivalent to \tfe{}
  for $e=0$. Bottom-right: Recovery of same set of aligned-spin eccentric events injected with \teo{} that
  include eccentricity,
  but recovered with the non-eccentric version of \teo{}. For all plots, the dashed line
  indicates the 90\% credible interval expected for a cumulative distribution
  drawn from 100 uniformly-distributed samples. All panels show recovery of
  the same injection set (i.e. all injection parameters are identical per event except for the waveform model),
  demonstrating that eccentric physics must be included in recovering eccentric waveforms regardless of
  waveform model.}
\label{fig:as_pp_plots}
\end{figure*}

We use a PP-plot to examine whether the signal parameters are recovered well
over the whole set of synthetic signals. A diagonal PP-plot indicates that the
distribution of the recovered signals matches the distribution of those
injected, as described in \secref{ss:synthpop}. For our aligned-spin eccentric
signals, we obtain a diagonal PP-plots shown in \figref{fig:as_pp_plots}; our
inference reliably recovers the source parameters, including eccentricity and
aligned-spin.


To perform the aligned-spin analysis, we use the Rotated Inspiral-Phase (RIP)
coordinates added in \cite{Wofford_2022} in the CIP step of our parameter
recovery. These coordinates are particularly useful in the low-mass regime.

PP-like plots also allow us to demonstrate conclusively just how
important eccentricity is when characterizing source properties.
\figref{fig:tfe_diag_as} shows that our parameter inference is unbiased for
\tfe{}.
Then we repeat the analysis described above, but require exactly quasicircular
orbits during inference: we recover eccentric non-spinning signals with a
non-eccentric, non-spinning waveform. Here we
use \texttt{TaylorF2}, which is the same as the \tfe{} model used above but
excludes eccentric corrections.

The same process is performed again using \teo{}. For this analysis we include
higher order modes up to $\ell_{\rm{max}}=4$ except for $m=0$ according to the
available subdominant modes for this waveform. \figref{fig:teo_diag_as} shows
that
our parameter inference is unbiased for this second waveform as well.
Then we once again repeat the analysis, but require quasicircular templates from
\teo{} during inference: we recover a  eccentric (non-spinning) signals made
with the eccentric version of \teo{} with the non-eccentric, non-spinning
version of \teo{}. 

\section{Conclusions}
We have demonstrated that RIFT can recover eccentricity in both the
non-spinning and aligned-spin cases using \tfe{} and \teo{} for low-mass binary
black holes. This study illustrates that it is crucial to include eccentricity
in parameter estimation for eccentric binary systems, otherwise recovery of
source parameters will be skewed.   Conversely, corroborating prior investigations, our study demonstrates that in low
mass binaries the eccentricity can be measured with exquisite precision, owing to the large number of cycles available
in these systems.
 Finally, we introduce a new parameter
to the RIFT pipeline to prepare for future parameter estimation systematics as
we await new discoveries in LIGO O4. This work also prepares the pipeline to
perform systematics studies for systems that include both eccentricity and
spin-precession as additional waveform models become available.



As noted in prior work, our study also demonstrates how waveform systematics can impact the recovery of parameters like
eccentricity; see, e.g., our Figure 4 for an example.  Evidently, precise measurements of eccentricity in low-mass
binary black holes may require further improvements in waveform models.  That said, in our study, contemprary models
seem to clearly
unambiguously indicate the impact of eccentricity, such that the presence of measurable eccentricity is more robustly
characterized than its precise value.  Given the astrophysical implications of the unambiguous identification of
eccentricity in a compact binary inspiral waveform, our investigation suggests that parameter inferences with
contemporary models provide a constructive avenue to identify eccentricity in low-mass binaries.

\begin{acknowledgements}
  ROS and KJW gratefully acknowledge support from NSF award PHY-2012057. ROS
  is acknowledges support from NSF PHY-2309172 and PHY-2207920. The authors
  are grateful for computational resources provided by the LIGO Laboratories
  at CIT, LHO, and LLO supported by National Science Foundation Grants
  PHY-0757058 and PHY-0823459 which are used for the parameter estimation runs.
  This material is based upon work supported by NSF’s LIGO Laboratory which
  is a major facility fully funded by the National Science Foundation.
\end{acknowledgements}

\bibliography{biblio,LIGO-publications}

\end{document}